\documentclass{iaus}
\usepackage{graphicx}

\usepackage{natbib} 
\bibpunct[,]{(}{)}{;}{a}{}{,} 
\newcommand{\SgrA}{Sgr~A$^\ast$}

\newcommand{\Msun}{\ensuremath{{\rm M}_{\odot}}}

\newcommand{\Rstar}{\ensuremath{r_{\ast}}}

\newcommand{\Rsun}{\ensuremath{{\rm R}_{\odot}}}

\newcommand{\kms}{\ensuremath{\mathrm{km\,s}^{-1}}}

\newcommand{\tcoll}{\ensuremath{t_{\rm coll}}}

\newcommand{\rem}[1]{}

\title[Dynamics of galactic nuclei] 
{Dynamics of galactic nuclei:\\Mass segregation and collisions}

\author[Freitag et al.]   
{Marc Freitag$^1$, James E.\ Dale$^2$, Ross P.\ Church$^{2,3}$ \break
\and Melvyn B.\ Davies$^2$%
}
\affiliation{$^1$Institute of Astronomy, Madingley road, Cambridge, CB3~0HA, UK \break email: freitag@ast.cam.ac.uk\\[\affilskip]
$^2$Lund Observatory, Box 43, SE~221Ð00, Lund, Sweden\\[\affilskip]
$^3$Centre for Stellar and Planetary Astrophysics, Monash University, Vic.\ 3800, Australia}

\pubyear{2008}
\volume{245}  
\pagerange{xxx--yyy}
\date{?? and in revised form ??}
\jname{Formation and Evolution of Galaxy Bulges}
\editors{M. Bureau, E. Athanassoula, \& B. Barbuy, eds.}

\begin{document}

\maketitle

\begin{abstract}
Massive black holes (MBHs) with a mass below $\sim 10^7\,{\rm
M}_\odot$ are likely to reside at the centre of dense stellar
nuclei shaped by 2-body relaxation, close interactions with
the MBH and direct collisions. In this contribution, we stress
the role of mass segregation of stellar-mass black holes into the
innermost tenths of a parsec and point to the importance of
hydrodynamical collisions between stars. At the Galactic centre, 
collisions must affect giant stars and some of the 
S-stars.
\end{abstract}

\firstsection 
\section{Introduction}

It is now firmly established that Sgr\,A$^*$, the weak radio source at
the centre of the Milky Way, is a massive black hole (MBH) with a mass
of $M_\bullet\simeq 4\times 10^{6}\,{\rm M}_\odot$. The cluster of
stars around it can be observed at resolutions as high as $\sim
0.002$\,pc in the near infra-red, allowing us to witness the unique
stellar dynamical processes at play in a galactic nucleus
\citep[e.g.,][and Sch\"odel, these
proceedings]{EisenhauerEtAl05,SchoedelEtAl07}.

A simple model for the stellar density within a few parsecs of
Sgr\,A$^*$, compatible with the observed stellar kinematics is given
by
\begin{equation}
\rho(R) \simeq \frac{3-\gamma}{4\pi} 
\frac{M_\bullet}{R_{\rm infl}^3}\left(\frac{R}{R_{\rm infl}}\right)^{-\gamma}
\simeq 5\times 10^{6}\,{\rm M}_\odot{\rm pc}^{-3}\,\left(\frac{R}{0.1\,{\rm pc}}\right)^{-1.5},
\label{eq.rho}
\end{equation}
where $R_{\rm infl}\simeq 2\,$pc is the influence radius and
$\gamma\simeq 1.5$. Well inside $R_{\rm infl}$, the (1--D) velocity
dispersion follows a Keplerian profile,
\begin{equation}
\sigma(R) \simeq \sqrt{\frac{1}{1+\gamma}\frac{GM_\bullet}{R}}
\simeq 260\,{\rm km\,s}^{-1}\,\left(\frac{R}{0.1\,{\rm pc}}\right)^{-0.5}.
\label{eq.sigma}
\end{equation}
Stars exchange energy and angular momentum through 2-body relaxation,
i.e.\ a multitude of 2-body scatterings, over a timescale
\citep{Spitzer87},
\begin{equation}
t_{\rm rlx}(R)=\frac{0.339}{\ln(M_\bullet/\langle m\rangle)}\frac{\sigma^3(R)}{G^2\langle m\rangle
\rho(R)}
\simeq 4\times 10^9\,{\rm yr},
\end{equation}
where we have a constant average mass $\langle m\rangle=1\,{\rm
M}_\odot$. We note that for $\gamma = 1.5$, the relaxation time is
approximately constant well within the sphere of influence. 

Assuming that the Sgr\,A$^\ast$ cluster is typical and that one can
scale to other galactic nuclei using the $M - \sigma$ relation in the
form $\sigma =
\sigma_{\rm MW} (M_\bullet/4\times 10^6\,{\rm M}_\odot)^{1/\beta}$ with
$\beta\approx4-5$
\citep{FM00,TremaineEtAl02}, one can estimate the relaxation 
time inside the sphere of influence in a galactic nucleus hosting an
MBH of mass $M_\bullet$ to be $t_{\rm rlx} \approx 4\times 10^9\,{\rm
yr}\,(M_\bullet/4\times 10^6\,{\rm M}_\odot)^{(2-3/\beta)}$. This
rough relation suggests that MBHs less massive than about $10^7\,{\rm
M}_\odot$ typically inhabit nuclei where relaxation plays an important
role. These are the MBHs which can produce gravitational waves
detectable by LISA \citep[e.g.,][]{Hogan07} and probably dominate the rate
of accretion flares due to tidal disruption of stars
\citep{WM04}.

We now consider the timescale for direct collisions between
stars. We assume two populations of stars, with stellar masses
$m_1$ and $m_2$, stellar radii $r_1$ and $r_2$, and Maxwellian
velocity distributions of (1--D) dispersions $\sigma_1$ and $\sigma_2$.
The average time for a star of type 1 to collide with a star of
type 2 is
\begin{equation}
\tcoll = \left[\sqrt{8\pi}n_2\sigma_{\rm rel}(r_1+r_2)^2
\left(1+\frac{G(m_1+m_2)}{\sigma_{\rm rel}^2(r_1+r_2)}\right)\right]^{-1},
\end{equation}
with $\sigma_{\rm rel}^2=\sigma_1^2+\sigma_2^2$ \citep{BT87}. At very
small distances from the MBH, using equations \ref{eq.rho} and
\ref{eq.sigma} and setting $m_2=\langle m \rangle=\Msun$ and $r_2=\Rsun$,
the collision time for a star of radius $r_1\equiv\Rstar$ is
$\tcoll(R) \approx 10\,{\rm Gyr} \left[(\Rstar+\Rsun)/2\,\Rsun\right]^{-2}({R}/{0.02\,{\rm pc}})^{2}$.
This suggests that a few thousand main-sequence (MS) stars should have suffered from
at least one collision in the Galactic centre.

\section{Mass segregation in galactic nuclei}

The main effect of 2-body relaxation in a galactic nucleus is to
produce mass segregation \citep*{FASK06,HA06b}. Stellar-mass black
holes (SBHs) are $10-20$ times more massive than the average star. In
our stellar-dynamical models of the Galactic nucleus, 2000 to 3000 of
them accumulate in the innermost 0.1\,pc over a timescale of
$3-5$\,Gyr. Without mass segregation, only 200 to 300 SBHs would be
expected there. The population within 1\,pc tallies $\sim 20\,000$.
They are swallowed by the MBH at a rate of about one per Myr. SBHs
dominate the mass density within $\sim 0.1$\,pc of the MBH. There they
form a cusp compatible with the profile predicted by
\citet{BW76}, $\rho(R) \propto R^{-\gamma}$ with $\gamma=1.75$, while
all lighter objects (including the neutron stars) form 
shallower cusps with $\gamma\simeq 1.3-1.5$.

Possible observational consequences of a population of
of SBHs in the Galactic centre are
mentioned by \citet{FASK06}. They include their appearance as X-ray
sources, accreting from a companion star or from interstellar gas
\citep{MunoEtAl05,DN07} and minute deflections of the trajectories 
of the S-stars \citep*{WMG05}.
\rem{Transient X-ray sources appear to be
overabundant in the central parsec of our Galaxy
\citep{MunoEtAl05}. Most of these sources are probably binaries in 
which a neutron star or a black hole accretes from a companion
star. Simulations of galactic nuclei including mass segregation and
binary interactions are required to determine whether these effects
can combine to explain the observed distribution. It has been
suggested by \citet{NS07} and \citet{DN07} that single BHs accreting
interstellar gas can be amongst these sources and may collectively
outshine the central MBH in some low-luminosity AGNs. A detection of
the dark cusp of BHs through the Newtonian retrograde precession it
induces on orbits of visible stars (such as the ``S-stars'') at the
Galactic centre seems unlikely, even with the next generation of
``extremely large telescopes'', but such facilities should make it
possible to observe 2-body gravitational deflections of the
trajectories of visible stars by BHs \citep{WMG05}.}  SBHs can spiral
into MBHs by emission of gravitational waves. If it is not perturbed
into a direct plunge or a wider non-inspiralling orbit before it
reaches a period of about $10^4$\,s an SBH can become a continuous
source for LISA, detectable at several Gpc (see review by
\citealt{IEMRIpaper07}). The segregation of SBHs is key to obtain
inspiral rates of observational interest ($\gtrsim 10^{-8}\,{\rm
yr}^{-1}$ per galaxy). It is also possible that LISA will detect a few
bursts of gravitational radiation emitted by SBHs at the pericentre of
very eccentric but long-period orbits around the Galactic MBH
\citep*{HFL07}.

\section{Collisions at the Galactic centre}

\begin{figure}
\centering
\resizebox{0.7\hsize}{!}{\includegraphics{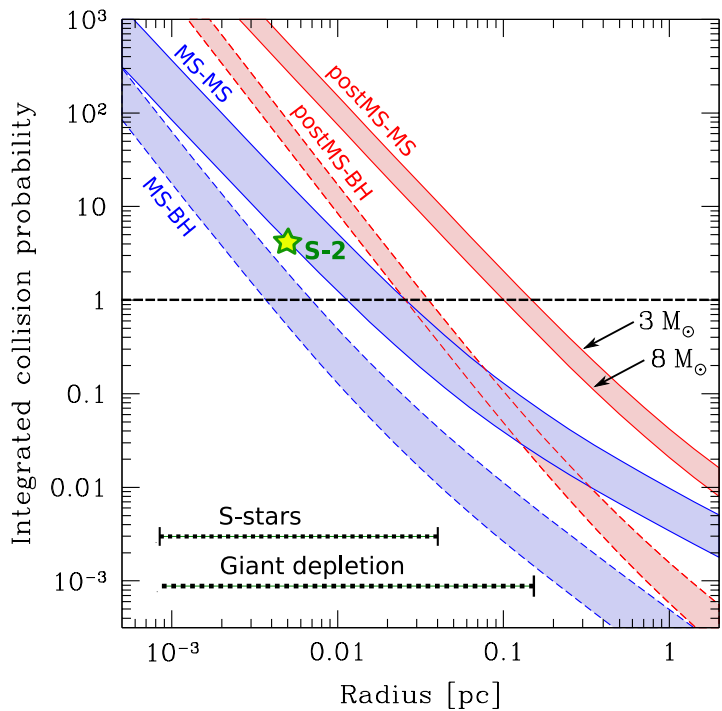}}
\caption[]{Collision probabilities at the Galactic centre. We plot the
probability for a star of 3 or $8\,\Msun$ to collide with a MS star or
a stellar-mass BH, integrated over the MS or post-MS phase of its
evolution, $P_{\rm coll}=\int t_{\rm coll}^{-1}dt$ (Dale et al., in
preparation). When $P_{\rm coll}$ is larger than one, it can be
interpreted as the expected number of collisions during this phase,
neglecting perturbations of the stellar evolution caused by previous
collisions. Stellar evolution models are used to determine
$\Rstar(t)$. We assume the field MS stars are $0.5\,\Msun$ objects and
follow a density profile $n(R)=n_0(R/0.1\,{\rm pc})^{-\gamma}$ with
$n_0\simeq10^7\,{\rm pc}^{-3}$ and $\gamma=1.4$. The BHs have a mass
of $10\,\Msun$, $n_0\simeq 5\times 10^5\,{\rm pc}^{-3}$ and
$\gamma=1.8$. These profiles are inspired by detailed stellar
dynamical models
\citep{FASK06}. We indicate the range of radii where depletion of
bright giants is observed, where the S-stars are, and an estimate of
the value of 
$P_{\rm coll}$ for S-2.\label{fig.Pcoll}}
\end{figure}

Collisions are unlikely to play an important role in the global dynamics of
the {\SgrA} cluster or to provide the Galactic MBH with
significant amounts of gas in comparison with tidal disruptions or
stellar winds
\citep{FASK06}. However collisions certainly occur in the Galactic 
centre, with potentially important observational consequences. This
point is conveyed by Fig.~\ref{fig.Pcoll} in which we plot an estimate
of the collision probability over the MS and post-MS lifetime of
massive stars. As possible impactors, we consider MS stars and
(mass-segregated) SBHs. One notices that, during their MS phase, many
S-stars must experience one or several collisions with MS stars
or SBHs.

\begin{figure}
\centering
\resizebox{0.56\hsize}{!}{\includegraphics{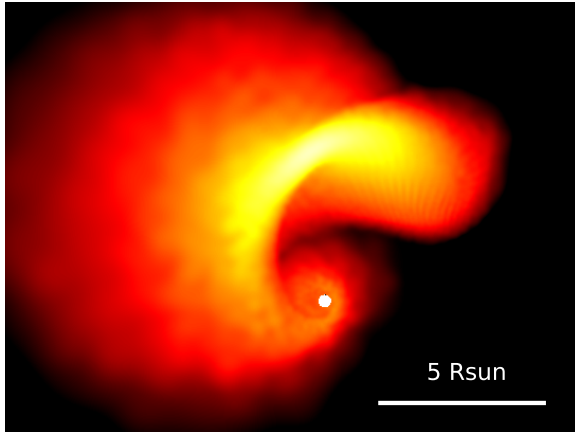}}
\caption[]{Snapshot of a SPH simulation of a collision between 
a $1\,\Msun$ MS star and a $10\,\Msun$ black hole (white dot).  The
relative velocity (at large separation) is $873\,\kms$ and the
pericentre distance assuming point-mass Keplerian trajectories is
$0.5\,\Rsun$ (Dale et al., in preparation).
\label{fig.coll}}
\end{figure}

To complement the work of \citet{FB05} on MS--MS collisions, we are
performing a large series of simulations of high-velocity encounters
between stars and SBHs, using the ``Smoothed Particle Hydrodynamics''
(SPH) method. A snapshot from one of our runs is shown in
Fig.~\ref{fig.coll}. First results suggest that SBHs can be damaging
even without a direct hit, by raising large tides in a MS star and
causing it to swell. Puffed-up stars would be much larger and more
fragile targets until they contract back to a normal size.

Furthermore, a star is much more likely to be struck once it has
become a giant. Nearly all stars within 0.1\,pc of the Galactic centre
must collide during their post-MS life. The population of bright
giants is observed to be significantly depleted within $\sim 0.15\,$pc
of the Galactic centre \citep{GTKKTG96}. We are investigating whether
collisions can be responsible, either by stripping the giants'
envelopes \citep{BD99} or by reducing the mass of MS stars, thus
preventing them from becoming giants.

\pagebreak 
\begin{acknowledgments}
The work of MF is founded through the STFC theory rolling grant to the
Institute of Astronomy in Cambridge. 
JED's work is supported by a stipend from the Wenner-Gren foundations.
RPC's work in Lund was funded by
a Swedish Institute Guest Scholarship. MBD is a Royal Swedish Academy
Research Fellow supported by a grant from the Knut and Alice
Wallenberg Foundation.
\end{acknowledgments}


\label{lastpage}

\end{document}